\documentclass[pre,twocolumn,showpacs,amsmath,amssymb]{revtex4}
\usepackage[dvipdfmx]{graphicx}
\usepackage{epstopdf}
\usepackage{dcolumn}
\usepackage{bm}
\usepackage{booktabs}

\begin{document}

\title{Critical nonequilibrium relaxation in the Swendsen-Wang algorithm 
in the Berezinsky-Kosterlitz-Thouless and weak first-order phase transitions}

\author{Yoshihiko Nonomura}
\email{nonomura.yoshihiko@nims.go.jp}
\affiliation{Computational Materials Science Unit, 
National Institute for Materials Science, Tsukuba, Ibaraki 305-0044, Japan} 

\author{Yusuke Tomita}
\email{ytomita@shibaura-it.ac.jp}
\affiliation{College of Engineering, Shibaura Institute of Technology, 
Saitama 337-8570, Japan} 

\begin{abstract}
Recently we showed that the critical nonequilibrium relaxation 
in the Swendsen-Wang algorithm is widely described by the 
stretched-exponential relaxation of physical quantities in the 
Ising or Heisenberg models. Here we make a similar analysis 
in the Berezinsky-Kosterlitz-Thouless phase transition in the 
two-dimensional (2D) XY model and in the first-order phase 
transition in the 2D $q=5$ Potts model, and find that these phase 
transitions are described by the simple exponential relaxation 
and power-law relaxation of physical quantities, respectively. 
We compare the relaxation behaviors of these phase transitions 
with those of the second-order phase transition in the 3D and 4D 
XY models and in the 2D $q$-state Potts models for $2 \le q \le 4$, 
and show that the species of phase transitions can be clearly 
characterized by the present analysis. We also compare the 
size dependence of relaxation behaviors of the first-order 
phase transition in the 2D $q=5$ and $6$ Potts models, 
and propose a quantitative criterion on ``weakness" 
of the first-order phase transition.
\end{abstract}


\maketitle

\section{Introduction}
The cluster algorithms have been proposed to overcome 
the critical slowing down. The dynamical critical exponent 
$z$($\approx 2$) in the local-update algorithms 
was reported to be reduced of 
one order by the cluster algorithms~\cite{SW,Wolff}. 
In numerical calculations, this exponent was evaluated 
from size dependence of the correlation time of equilibrium 
autocorrelation functions of physical quantities at the critical 
temperature, $\tau(L) \sim L^{z}$. It was gradually revealed 
that to distinguish power-law behavior with such a small 
exponent from logarithmic behavior, $\tau(L) \sim \log L$, 
is quite difficult~\cite{Heermann,Baillie}. 

Then, larger-scale calculations based on nonequilibrium 
relaxation (NER)~\cite{NERrev} showed~\cite{Gunduc,Du} 
that a critical relaxation faster than power law (symbolically 
represented as $z=0$) might exist in the cluster algorithms. 
Quite recently, the present authors showed that the explicit 
form of such a nontrivial critical relaxation is given by the 
stretched-exponential time dependence, initially in the 
two-dimensional (2D) Ising model in the Wolff algorithm from 
the perfectly-ordered state~\cite{Nonomura14}, and then 
in the 3D Heisenberg model in the Swendsen-Wang (SW) 
algorithm from the perfectly-disordered state~\cite{Nono-Hei}.

In the standard NER based on the local-update algorithms, 
a physical quantity of large enough size shows a power-law 
critical relaxation $\langle Q \rangle \sim t^{\mp \psi /(z \nu)}$, 
with the critical exponent $\psi$ of the quantity $Q$ 
and $\nu$ of the correlation length for decaying ($-$) 
or ordering ($+$) processes. The critical temperature 
can be evaluated from the power-law behavior 
(or linearity of the $\log$-$\log$ plot), and the critical 
exponents can be obtained from the mixed exponents 
of some quantities and scaling relations. 

Once the functional form of the critical relaxation 
in the cluster NER is established, analysis similar 
to that in the standard NER becomes possible, and 
the critical temperature can be evaluated from the 
stretched-exponential behavior~\cite{Nono-Hei}. 
However, the critical exponents are not included in 
the stretched-exponential time dependence, and 
such information can be obtained only from the 
nonequilibrium-to-equilibrium scaling~\cite{Nonomura14}, 
a kind of finite-size scaling which couples the short-time 
critical relaxation and equilibrium size dependence. 
Although this procedure ensures the nontrivial 
stretched-exponential behavior in the cluster NER, 
necessity of multiple sizes is not so favorable as 
the standard NER, and necessity 
of relaxation data until the vicinity of  equilibrium 
is also not favorable because numerical merit 
in comparison with conventional equilibrium 
simulations becomes obscure.

In the present article we numerically show that 
the essential merit of the cluster NER is exhibited 
in the Berezinsky-Kosterlitz-Thouless (BKT) and 
weak first-order phase transitions. 
Critical relaxation behaviors in the cluster NER are 
qualitatively different in the BKT or weak first-order 
phase transitions and in the second-order phase 
transition, and nontrivial physical results can be 
obtained only from initial relaxation behaviors.

The outline of the present article is as follows. In section II, 
the basic scaling form of magnetization in the critical 
relaxation with respect to time and size is exhibited. 
In section III, we calculate this time dependence in 
the 2D, 3D and 4D XY models, and show that the 
peculiar behavior in 2D characterizes the BKT phase 
transition. In section IV, critical relaxation in the 2D 
$q$-sate Potts models are compared with each other 
for $2 \le q \le 8$, and the qualitative difference between 
$2 \le q \le 4$ (second-order phase transition) and 
$q \ge 5$ (first-order phase transition) is displayed. 
The size dependence of the ordering process in the 
first-order phase transition is investigated for $q=5$ 
and $6$, and weak-first-order nature for $q=5$ is 
visualized. In section V, numerical results in the 
present article are discussed. The meaning of the universal 
stretched-exponential exponent observed in the 3D and 
4D XY models is considered, and a quantitative criterion on 
``weakness" of the first-order phase transition is proposed. 
In section VI, the above descriptions are summarized. 

\section{Formulation}
In the present article the 2D Potts models are analyzed 
with the SW algorithm~\cite{SW}, and the XY models are 
treated with the ``embedded-Ising-spin" scheme proposed 
by Wolff~\cite{Wolff}, although spin clusters are generated 
in the whole system after the SW algorithm, not after Wolff's 
single-cluster scheme (the so-called Wolff algorithm)~\cite{Wolff}. 
Since the purpose of the present article is to exhibit nontrivial 
critical relaxation in the BKT and weak first-order phase transitions, 
here we do not intend to evaluate the critical temperature 
$T_{\rm c}$ precisely. Instead, we use the exact value 
$T_{\rm c}=1/\log ( 1 + \sqrt{q} )$ [$J/k_{\rm B}$] 
in the 2D $q$-state Potts model on a square lattice with 
coupling $J$. In the 3D and 4D XY models, a rather rough 
evaluation of $T_{\rm c}$ is enough for the estimation of 
the stretched-exponential exponent. In the 2D XY model, 
the BKT phase transition results in a large finite-size effect, 
and size dependence of $T_{\rm c}$ is crucial. 
In this case a recent large-scale simulation by 
Komura and Okabe \cite{Komura} is referred to. 

In the XY models, we concentrate on the nonequilibrium 
ordering from the perfectly-disordered state, because 
early-time deviation from the stretched-exponential 
form is much smaller than that of the decay from the 
perfectly-ordered state in vector spin models~\cite{Nono-Hei}. 
The absolute value of the magnetization behaves as 
\begin{equation}
\label{se-dis}
\langle |m(t,L)| \rangle \sim L^{-d/2} \exp \left[ + ( t / \tau_{m} )^{\sigma} \right] \ 
(0<\sigma <1),
\end{equation}
with the spatial dimension $d$, the ``relaxation time" $\tau_{Q}$ depending 
on a quantity $Q$ and the exponent $\sigma$ independent of quantities. 
Here the size dependence originates from the normalized 
random-walk growth of clusters. In order to evaluate the 
exponent $\sigma$ simply, we utilize the double-log plot, 
\begin{equation}
\log \left[ \log ( \langle |m(t,L)| \rangle / C(L) ) \right] \sim \sigma \log ( t /\tau_{m}),
\end{equation}
which is directly derived from Eq.\ (\ref{se-dis}). Here $C(L)$ 
stands for the coefficient of rhs of Eq.\ (\ref{se-dis}) including 
size dependence. The unknown coefficient $C(L)$ can be 
determined by aligning the early-time data on a straight line. 

In the 2D Potts models, we calculate both decay 
from the perfectly-ordered state and ordering from 
the perfectly-disordered state, because the relaxation 
process becomes slower and slower as the number 
of state $q$ increases, and whether the system 
is still trapped by the pure ordered or disordered 
stable states or already arrives at equilibrium 
(coexistence of the two stable states) in the 
first-order phase transition can be distinguished 
only by comparing these two relaxation processes. 
Weakness of the first-order phase transition is 
mainly characterized by the ordering process 
from the perfectly-disordered state. 

\section{Numerical results in the XY models: 
BKT versus second-order phase transitions}
\begin{figure}
\includegraphics[width=88mm]{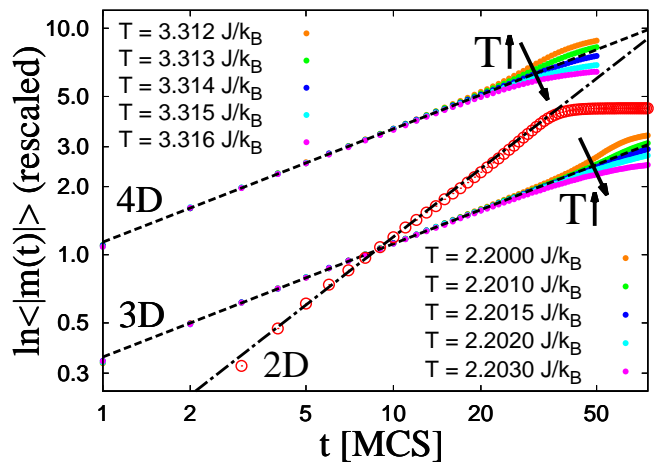}
\caption{(Color online) Double-log plot of time dependence 
of absolute value of magnetization during ordering process 
from the perfectly-disordered state in the vicinity of $T_{\rm c}$ 
for the 3D XY model for $L=400$ and 4D XY model for $L=100$ 
with the dashed lines corresponding to $\sigma=1/2$. 
The data for the 2D XY model for $L=8000$ at 
$T=0.90 J/k_{\rm B}$ are also plotted by open circles 
with the chain line corresponding to $\sigma=1$.
}
\label{fig-XY-2nd}
\end{figure}
First, a double-log plot of the absolute value of magnetization 
obtained from ordering from the perfectly-disordered state is 
displayed in Fig.\ \ref{fig-XY-2nd} for the 3D XY model for 
$L=400$ and for the 4D XY model for $L=100$. In the 3D XY model, 
the tangent of early-time behavior is almost unchanged for varying 
the temperature for $2.200 J/k_{\rm B} \le T \le 2.203 J/k_{\rm B}$ 
with $\tau_{m} \approx 2.00$, $C(L=400) \approx 9.5 \times 10^{-5}$ 
and $\sigma \approx 0.5$. Similar behavior is observed for the 
4D XY model for $3.312 J/k_{\rm B} \le T \le 3.316 J/k_{\rm B}$ 
with $\tau_{m} \approx 1.74$, $C(L=100) \approx 7.6 \times 10^{-5}$ 
and $\sigma \approx 0.5$. The exponent $\sigma \approx 0.5$ 
is also observed in the 3D Heisenberg model~\cite{Nono-Hei}, 
and this exponent might be universal in vector spin models. 

Then, the absolute value of magnetization in the 2D XY model 
for $L=8000$ (total number of spins is the same as that in the 
3D XY model for $L=400$) is also plotted in the same figure 
(open circles). Here the data at $T=0.90 J/k_{\rm B}$ are 
exhibited following $T_{\rm KT} \approx 0.8992 J/k_{\rm B}$ 
for $L=8192$~\cite{Komura}. Aside from initial a few 
Monte Carlo steps (MCS), these data are well on 
a straight (chain) line with $\tau_{m} \approx 5.00$, 
$C(L=8000) \approx 2.05 \times 10^{-4}$ and $\sigma \approx 1.0$. 
Such time dependence is clearly different from those in the 3D and 4D 
XY models characterized by $\sigma \approx 0.5$. 

\begin{figure}
\includegraphics[width=88mm]{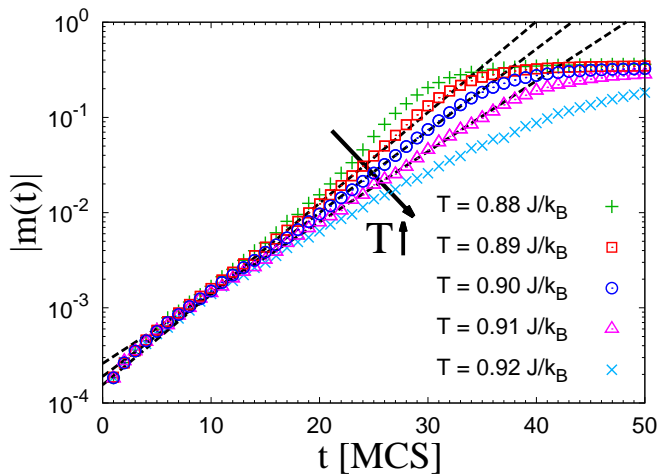}
\caption{(Color online) Semilog plot of time dependence of 
absolute value of magnetization during ordering process 
from the perfectly-disordered state in the 2D XY model 
for $L=8000$ for $T=0.88 J/k_{\rm B}$ (crosses), 
$0.89 J/k_{\rm B}$ (squares), $0.90 J/k_{\rm B}$ (circles), 
$0.91 J/k_{\rm B}$ (triangles) and $0.92 J/k_{\rm B}$ (X-marks), 
and the linearity of the data is emphasized by dashed lines.
}
\label{fig-XY-all}
\end{figure}

The case $\sigma=1$ is nothing but the 
simple-exponential time dependence. In such 
a case a naive semilog plot is more convenient 
and reliable than the double-log plot, because 
the unknown parameter $C(L)$ does not appear. 
The data for $0.88 J/k_{\rm B} \le T \le 0.92 J/k_{\rm B}$ 
are plotted in Fig.\ \ref{fig-XY-all}, and at least the data for 
$T=0.89 J/k_{\rm B}$ (squares), $0.90 J/k_{\rm B}$ (circles) 
and $0.91 J/k_{\rm B}$ (triangles) seem consistent with 
the simple-exponential time dependence (dashed lines). 
The nontrivial simple-exponential behavior in two dimensions 
suggested in Fig.\ \ref{fig-XY-2nd} is confirmed, and such a 
breakdown of the universal relaxation behavior in XY models 
only in two dimensions would be the consequence of the 
BKT phase transition, which occurs only in two dimensions. 

\section{Numerical results in the 2D $q$-state 
Potts models for various $q$}
\subsection{Comparison between the second-order 
($2 \le q \le 4$) and first-order ($q \ge 5$) phase transitions}
\begin{figure}[t]
\includegraphics[width=88mm]{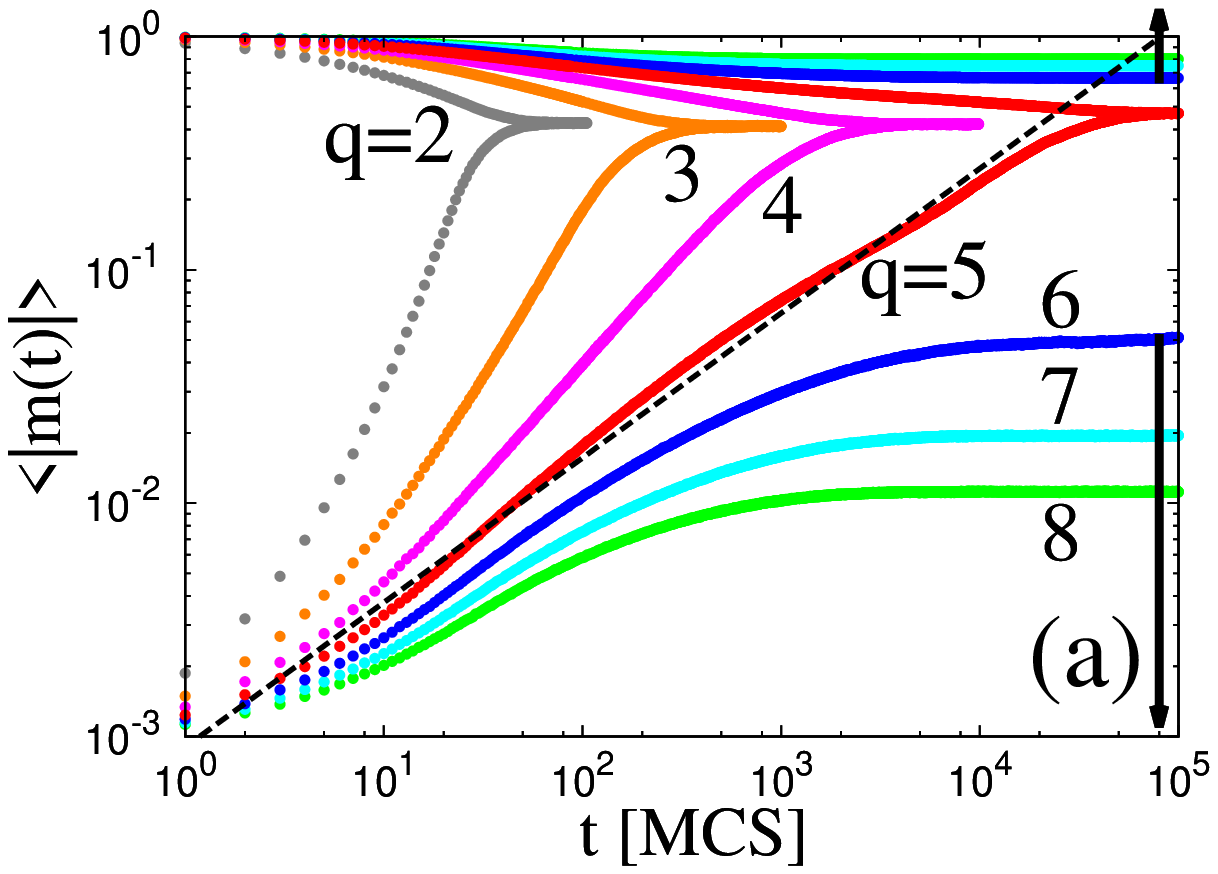}
\includegraphics[width=88mm]{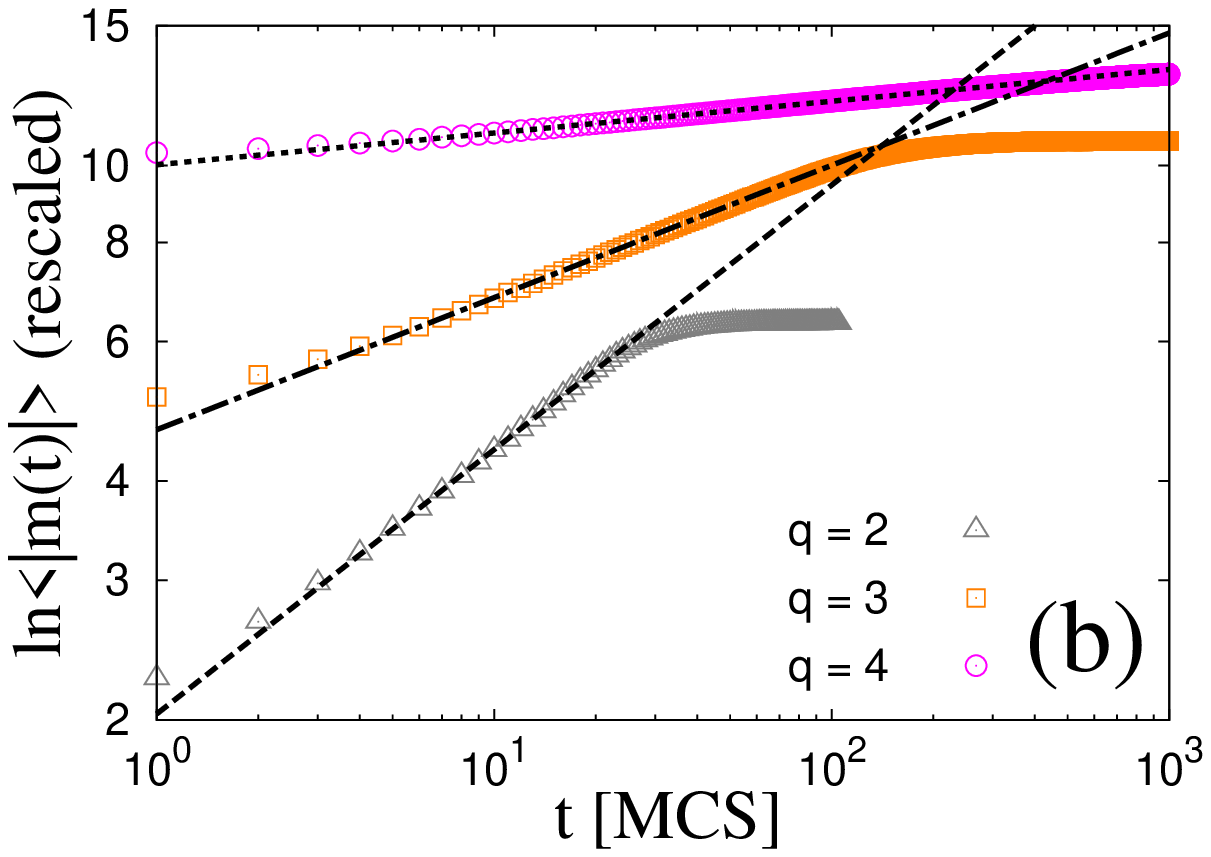}
\caption{(Color online) (a) Log-log plot of time dependence of 
absolute value of magnetization during the decaying process 
from the perfectly-ordered state and the ordering process from 
the perfectly-disordered state at the exact $T_{\rm c}$ in the 
2D $q$-state Potts models for $2 \le q \le 8$ and $L=1000$. 
The stretched-exponential behavior is observed in the 
second-order phase transition for $2 \le q \le 4$, and 
the exponent $\sigma$ decreases from $\sigma=1/3$ 
at $q=2$ (Ising case) as $q$ increases. 
Almost power-law behavior (drawn by the dashed line) 
is observed at $q=5$ for this system size. 
Trapping behavior to pure stable states indicates 
the typical first-order phase transition for $q \ge 6$, 
where the magnetization gap increases as $q$ 
increases as shown by arrows. 
(b) Double-log plot of the above data for $2 \le q \le 4$, 
i.e.\ the data for $q=2$ (triangles) with the dashed 
line corresponding to $\sigma=1/3$, the data for 
$q=3$ (squares) with the chain line representing 
$\sigma=1/6$, and the data for $q=4$ (circles) 
with the dotted line assuming $\sigma=0.04$.
}
\label{fig-Potts-all}
\end{figure}
Decaying behavior from the perfectly-ordered state and 
ordering behavior from the perfectly-disordered state in 
the 2D $q$-state Potts models for $2 \le q \le 8$ and 
$L=1000$ are summarized in Fig.\ \ref{fig-Potts-all}(a), 
namely the log-log plot of time dependence of the 
absolute value of magnetization at the exact $T_{\rm c}$. 
This model exhibits the second-order phase transition for 
$2 \le q \le 4$, which results in the stretched-exponential 
time dependence in the critical cluster NER. 
As shown in Fig.\ \ref{fig-Potts-all}(b), equilibration becomes 
slower as the number of state $q$ increases, and the exponent 
$\sigma$ decreases from $\sigma=1/3$ (dashed line) 
at $q=2$ (Ising case, triangles)~\cite{Nonomura14}. 
At $q=3$ (squares), this exponent seems consistent 
with $\sigma=1/6$ (chain line). At $q=4$ (circles), 
the exponent is much reduced and a preliminary 
estimate from these data is $\sigma \approx 0.04$. 
In this figure we draw a dotted guide line with $\sigma=0.04$. 
Note that a few initial-time data points are not on straight lines, 
and the deviation is much larger than the cases in the 3D 
and 4D XY models as displayed in Fig.\ \ref{fig-XY-2nd}. 
Since Potts variables take discrete values, initial clusters are 
generated discontinuously in this model, and deviation in the 
initial relaxation process is inevitable. Further investigations 
for $q=3$ and $4$ will be given elsewhere.

This model shows the first-order phase transition 
for $q \ge 5$, and the trapping behavior to pure 
astable states, which is a typical sign of the first-order 
phase transition, is observed for $q \ge 6$, and the 
magnetization gap increases as $q$ increases 
(drawn by arrows)~\cite{Baxter82}. 
This behavior has already been reported in 
NER based on the local-update algorithm 
at $q=20$~\cite{Ozeki03}, and also based on 
the SW algorithm at $q=10$~\cite{Kasono}. 
In these cases initial relaxation processes 
are described by the power-law relaxation, 
and then saturate into pure stable states. 
At $q=5$ for $L=1000$, such behavior is 
replaced by convergence to equilibrium 
(coexistence of two stable states) almost 
in a power law (drawn by the dashed line). 

\subsection{Comparison of the size dependence in the 
first-order phase transition for $q=5$ and $q=6$}
\begin{figure}[t]
\includegraphics[width=88mm]{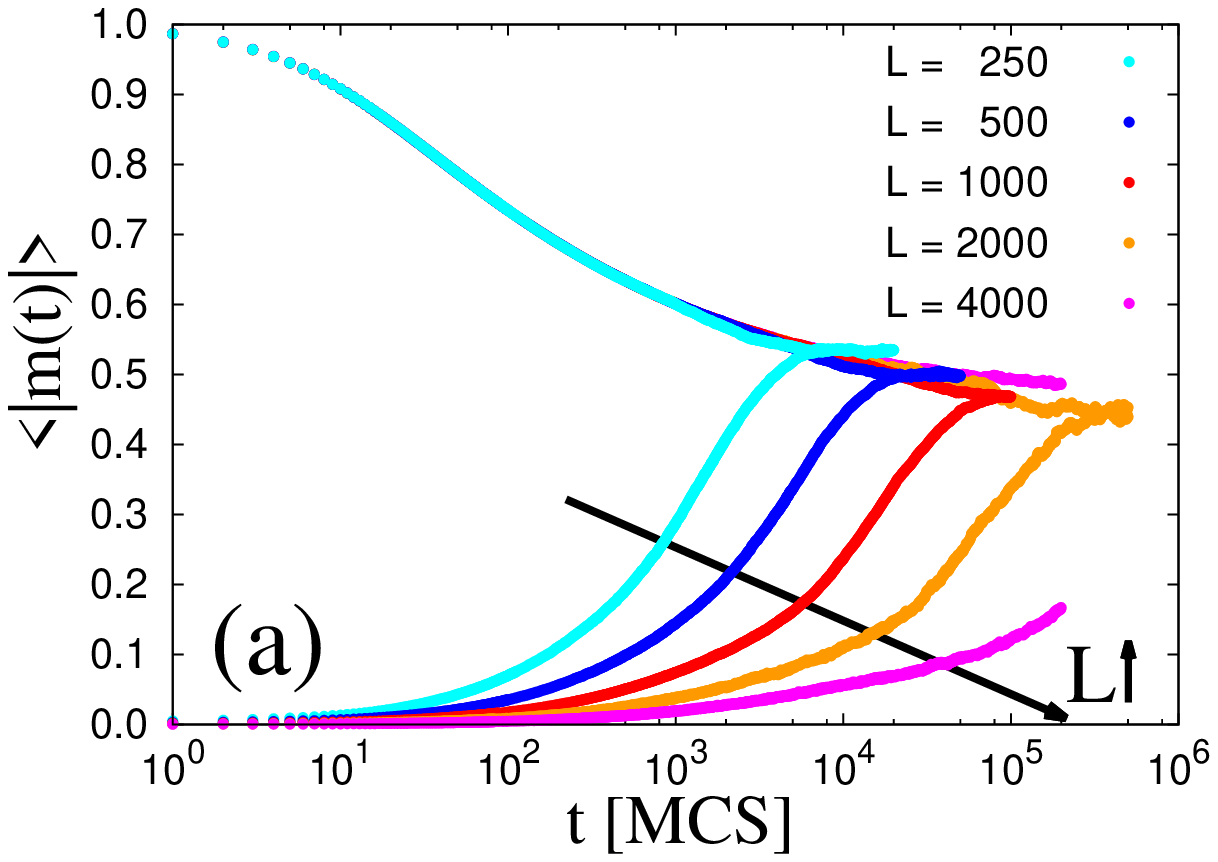}
\includegraphics[width=88mm]{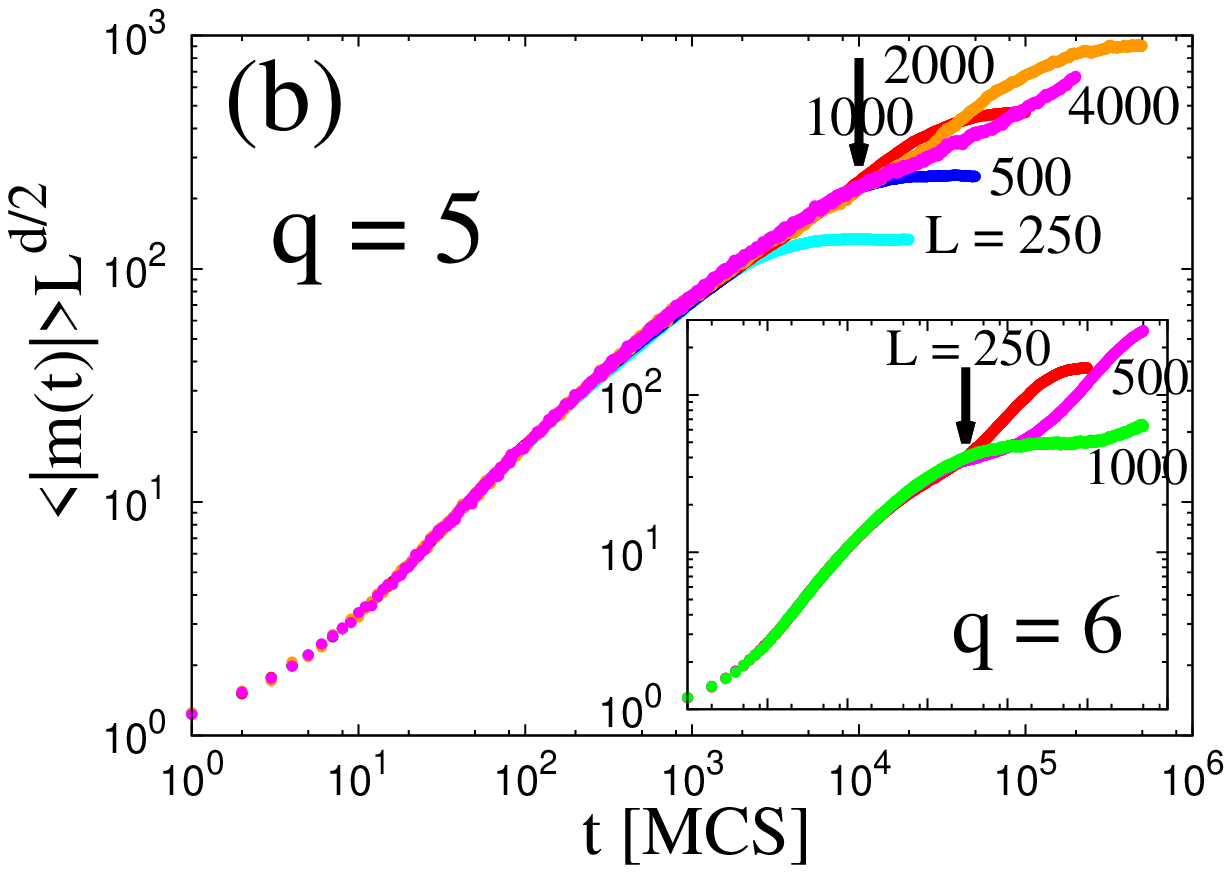}
\caption{(Color online) Time dependence of absolute value of 
magnetization at the exact $T_{\rm c}$ for various system sizes 
(a) in the 2D $q=5$ Potts model during decaying process from 
the perfectly-ordered state and ordering process from the 
perfectly-disordered state in an inverse semilog plot, and 
(b) in the 2D $q=5$ and $6$ (inset) Potts models from the 
perfectly-disordered state adjusted by the random-walk 
growth of clusters $\sim L^{-d/2}$ in a log-log plot. 
All the data are located on a single curve up to 
the onset of trapping behavior (arrows).
}
\label{fig-Potts-RW}
\end{figure}

In order to understand the origin of such nontrivial 
behavior in the first-order phase transition at $q=5$, 
size dependence of the data is investigated. 
The data corresponding to Fig.\ \ref{fig-Potts-all}(a) 
for various system sizes at $q=5$ are plotted in 
Fig.\ \ref{fig-Potts-RW}(a) in an inverse semilog 
plot (the $x$-axis is drawn in a log scale), and the 
data for $250 \le L \le 2000$ converge to equilibrium 
up to $5 \times 10^{5}$ MCS. In this figure the data 
in the decaying process are on a single curve until 
the system in each size approaches the vicinity of 
equilibrium. The equilibrium value of magnetization 
decreases systematically as the system size increases. 
Note that fluctuating behavior for $L=2000$ is because of 
the limited number of averaged random-number sequences. 
Since the relaxation time until equilibrium seems 
proportional to $L^{d}$ in this case, numerical 
efforts are proportional to $L^{2d}$ together with 
trivial increase proportional to volume $L^{d}$. 

In Fig.\  \ref{fig-Potts-RW}(b), the data in the 
ordering process from the perfectly-disordered state 
at $q=5$ are plotted in a log-log scale by taking the 
size dependence of random-walk growth (\ref{se-dis}) 
into account. The data for various system sizes are 
on a single curve after adjusting this $\sim L^{-d/2}$ 
size dependence up to the onset of 
trapping behavior indicated by the arrow. Such behavior 
becomes clearer in comparison with the data at $q=6$ 
for smaller sizes in the inset. The apparent trapping 
behavior for $L=1000$ is not observed for smaller sizes, 
and the waving behaviors for $L=250$ and $500$ at 
$q=6$ seem comparable to those for $L=1000$ and 
$4000$ at $q=5$, respectively. This correspondence 
suggests that the clear trapping behavior may not be 
observed at $q=5$ even for $L \sim 1 \times 10^{4}$, 
which is beyond reach numerically at present.

\section{Discussion}
Simple-exponential critical relaxation in the 2D XY model is 
quite nontrivial, because such behavior is usually observed 
above $T_{\rm c}$ (from the perfectly-ordered state) or 
below $T_{\rm c}$ (from the perfectly-disordered state). 
Such anomalous critical relaxation looks suitable for the 
anomalous BKT phase transition specific to two dimensions. 
Actually, equivalence of the exponent $\sigma$ of the 
stretched-exponential behavior in the 3D and 4D XY 
models is equally nontrivial, because the upper critical 
dimension $d_{\rm c}$ of the XY model is four; 
that is, dynamical behaviors in the cluster NER might be 
different from the mean-field one even above $d_{\rm c}$. 
Further investigations on this topic in the Ising models 
are now in progress~\cite{Nono-letter}.

The waving behavior in the ordering process of magnetization 
in the 2D $q=5$ Potts model could be regarded as the sign of 
trapping behavior to pure stable states, which is characteristic 
to the first-order phase transition. However, the onset of this 
behavior coincides with that of the saturation behavior toward 
equilibrium for $L=500$, as shown in Fig.~\ref{fig-Potts-RW}(b). 
That is, no explicit behaviors to represent the first-order 
phase transition can be observed for $L \le 500$. 
Such a critical system size can be much larger when 
the first-order phase transition is much weaker, and 
direct numerical evidence of the first-order phase 
transition becomes beyond reach in such a case. 

It would rather be a quantitative criterion of the ``weak 
first-order phase transition" that no signs of trapping 
behavior to pure stable states are observed even in 
the numerically-accessible maximum system size. 
Nevertheless, power-law behavior in the cluster NER 
exists even in such a case. Analytic treatment of the cluster 
NER is possible in the infinite-range model, and the study 
on the infinite-range $q=3$ Potts model exhibits~\cite{Tomita} 
similar power-law behavior in the first-order phase transition. 

\section{Summary}
In the present article the BKT phase transition in the 2D XY model and 
the (weak) first-order phase transition in the 2D $q=5$ Potts model are 
analyzed with the nonequilibrium relaxation (NER) in the Swendsen-Wang 
algorithm. Ordering process shows the simple-exponential critical 
relaxation in the former, and it shows the power-law critical relaxation 
in the latter. Both behaviors are qualitatively different from those 
in the second-order phase transition such as in the 3D and 4D 
XY models or in the 2D Potts model for $2 \le q \le 4$, 
which show the stretched-exponential critical relaxation. 
These behaviors can be regarded as the essential merit of the 
cluster NER, because nontrivial physical results can be obtained 
only from initial relaxation behaviors in these cases similarly to 
those in the standard NER.

The simple-exponential relaxation in the 2D XY model corresponds to 
$\sigma=1$ in the stretched-exponential time dependence (\ref{se-dis}), 
and such relaxation is usually observed above or below $T_{\rm c}$. 
The critical relaxation in the 3D and 4D XY models is characterized 
by the exponent $\sigma=1/2$ (such universality of the exponent 
$\sigma$ independent of the spatial dimension is also nontrivial), 
and this large deviation of $\sigma$ ensures anomaly in the 
BKT phase transition. The first-order phase transition in the 
2D $q \ge 6$ Potts models is characterized by the trapping 
behavior to pure stable states in dynamical processes, 
which is not observed in the 2D $q=5$ Potts model 
within the numerically accessible system sizes. 
Size dependence in the first-order phase transition is 
clarified by a kind of finite-size scaling analysis including 
the random-walk growth of spin clusters, and a quantitative 
criterion on the ``weak first-order phase transition" is proposed.
\bigskip
\par
\section*{Acknowledgments}
The random-number generator MT19937~\cite{MT} was used for 
numerical calculations. Most calculations were  performed on the 
Numerical Materials Simulator at National Institute for Materials Science.

\end{document}